\documentstyle[11pt,psfig,twoside]{article}

\begin{document}

\newcommand{\dd}{\,{\rm d}}
\newcommand{\ie}{{\it i.e.},\,}
\newcommand{\etal}{{\it et al.\ }}
\newcommand{\eg}{{\it e.g.},\,}
\newcommand{\cf}{{\it cf.\ }}
\newcommand{\vs}{{\it vs.\ }}
\newcommand{\zdot}{\makebox[0pt][l]{.}}
\newcommand{\up}[1]{\ifmmode^{\rm #1}\else$^{\rm #1}$\fi}
\newcommand{\dn}[1]{\ifmmode_{\rm #1}\else$_{\rm #1}$\fi}
\newcommand{\upd}{\up{d}}
\newcommand{\uph}{\up{h}}
\newcommand{\upm}{\up{m}}
\newcommand{\ups}{\up{s}}
\newcommand{\arcd}{\ifmmode^{\circ}\else$^{\circ}$\fi}
\newcommand{\arcm}{\ifmmode{'}\else$'$\fi}
\newcommand{\arcs}{\ifmmode{''}\else$''$\fi}
\newcommand{\MS}{{\rm M}\ifmmode_{\odot}\else$_{\odot}$\fi}
\newcommand{\RS}{{\rm R}\ifmmode_{\odot}\else$_{\odot}$\fi}
\newcommand{\LS}{{\rm L}\ifmmode_{\odot}\else$_{\odot}$\fi}

\newcommand{\Abstract}[2]{{\footnotesize\begin{center}ABSTRACT\end{center}
\vspace{1mm}\par#1\par
\noindent
{~}{\it #2}}}

\newcommand{\TabCap}[2]{\begin{center}\parbox[t]{#1}{\begin{center}
  \small {\spaceskip 2pt plus 1pt minus 1pt T a b l e}
  \refstepcounter{table}\thetable \\[2mm]
  \footnotesize #2 \end{center}}\end{center}}

\newcommand{\TableSep}[2]{\begin{table}[p]\vspace{#1}
\TabCap{#2}\end{table}}

\newcommand{\FigCap}[1]{\footnotesize\par\noindent Fig.\  %
  \refstepcounter{figure}\thefigure. #1\par}

\newcommand{\TableFont}{\footnotesize}
\newcommand{\TableFontIt}{\ttit}
\newcommand{\SetTableFont}[1]{\renewcommand{\TableFont}{#1}}

\newcommand{\MakeTable}[4]{\begin{table}[htb]\TabCap{#2}{#3}
  \begin{center} \TableFont \begin{tabular}{#1} #4 
  \end{tabular}\end{center}\end{table}}

\newcommand{\MakeTableSep}[4]{\begin{table}[p]\TabCap{#2}{#3}
  \begin{center} \TableFont \begin{tabular}{#1} #4 
  \end{tabular}\end{center}\end{table}}

\newenvironment{references}%
{
\footnotesize \frenchspacing
\renewcommand{\thesection}{}
\renewcommand{\in}{{\rm in }}
\renewcommand{\AA}{Astron.\ Astrophys.}
\newcommand{\AAS}{Astron.~Astrophys.~Suppl.~Ser.}
\newcommand{\ApJ}{Astrophys.\ J.}
\newcommand{\ApJS}{Astrophys.\ J.~Suppl.~Ser.}
\newcommand{\ApJL}{Astrophys.\ J.~Letters}
\newcommand{\AJ}{Astron.\ J.}
\newcommand{\IBVS}{IBVS}
\newcommand{\PASP}{P.A.S.P.}
\newcommand{\Acta}{Acta Astron.}
\newcommand{\MNRAS}{MNRAS}
\renewcommand{\and}{{\rm and }}
\section{{\rm REFERENCES}}
\sloppy \hyphenpenalty10000
\begin{list}{}{\leftmargin1cm\listparindent-1cm
\itemindent\listparindent\parsep0pt\itemsep0pt}}%
{\end{list}\vspace{2mm}}

\def\TYLDA{~}
\newlength{\DW}
\settowidth{\DW}{0}
\newcommand{\dw}{\hspace{\DW}}

\newcommand{\refitem}[5]{\item[]{#1} #2%
\def\REFARG{#3}\ifx\REFARG\TYLDA\else, {\it#3}\fi
\def\REFARG{#4}\ifx\REFARG\TYLDA\else, {\bf#4}\fi
\def\REFARG{#5}\ifx\REFARG\TYLDA\else, {#5}\fi.}

\newcommand{\Section}[1]{\section{#1}}
\newcommand{\Subsection}[1]{\subsection{#1}}
\newcommand{\Acknow}[1]{\par\vspace{5mm}{\bf Acknowledgments.} #1}
\pagestyle{myheadings}

\def\thefootnote{\fnsymbol{footnote}}

\begin{center}
{\Large\bf Variable Stars in the Field of Young Open Cluster NGC~581}

\vskip1cm

{\bf{\L}. ~~W~y~r~z~y~k~o~w~s~k~i$^1$,~~ G.~~ 
P~i~e~t~r~z~y~{\'n}~s~k~i$^{1,2}$~~ \\ and~~ O.~~ S~z~e~w~c~z~y~k$^1$}

\vskip3mm

{$^1$Warsaw University Observatory, Al.~Ujazdowskie~4, 00-478~Warszawa, Poland\\
email: (wyrzykow,szewczyk)@astrouw.edu.pl\\
$^2$Universidad de Concepci{\'o}n, Departamento de Fisica,
Casilla~160-C, Concepci{\'o}n, Chile\\
email: pietrzyn@hubble.cfm.udec.cl}
\end{center}

\Abstract{We present results of the search for variable stars in the field of 
young open cluster NGC~581. Based on 19 nights of observations, 6 new variable 
stars were discovered. Two of them turned out to be eclipsing binary systems. 
Another two detected variable stars are most probably of $\gamma$~Dor type. 
During our observations one of the known Be stars located in our field of view 
showed irregular variations of brightness, typical for this class of stars. 
The sixth variable star is a pulsating red giant.}{~}

{Stars: variables: general -- open clusters and associations: individual: NGC~581} 

\Section{Introduction}
Searches for variable stars in star clusters are very useful, because both 
variable stars and clusters are important sources of information on 
fundamental problems of stellar astrophysics. Variable stars which belong to 
cluster are especially well suited for studies of stellar dynamics and 
evolution. From their observations also  many information on cluster age, 
distance and chemical composition may be obtained. 

Our search is part of the long term project, aiming at systematic search for 
variable stars in open clusters initiated at the Warsaw University 
Observatory. Observations were conducted with both: 0.6~m telescope located 
at the Ostrowik Station near Warsaw and 1.3~m Warsaw telescope, located in 
the Las Campanas Observatory (LCO), Chile in the course of the OGLE project 
(Udalski, Kubiak and Szyma{\'n}ski 1997). At the Ostrowik Observatory four 
Northern Hemisphere clusters were monitored: NGC~654, NGC~663, IC~4996, 
NGC~659 (Pietrzy{\'n}ski 1996abcd, 1997, Pietrzy{\'n}ski 2001) and in total 
21 variable stars were discovered. Southern Hemisphere clusters were monitored 
from the LCO observatory and 59 new variable stars were discovered in the 
fields of two clusters: NGC~5999 and NGC~5381 (Pietrzy{\'n}ski \etal 1997, 1998). 
In this paper we present results of the  search for variable stars in the 
field of another, Northern Hemisphere open cluster: NGC~581. 

The cluster was cataloged already in XIX century by Messier (1850) with number 
103. Later it was observed by Alter (1940), who first derived its distance. 
Since then NGC~581 has been a subject of several investigations, including 
photographic {\it UBV} studies of Hoag \etal (1961), McCouskey and Houk 
(1964), Moffat (1972), Sagar and Joshi (1978), and {\it RGU} study by Steppe 
(1974). Recently, Phelps and Janes (1994), based on precise {\it BV} CCD 
photometry, derived the distance and the age of this cluster of about 2690~pc 
and 22~Myr, respectively. 

\Section{Observations}
Presented data were collected in 1999 during 19 nights at the Ostrowik Station 
of the  Warsaw University Observatory. The 0.6~m Zeiss reflector equipped  
with ${512\times512}$ pixel CCD detector was used. The scale of 0.74 
arcsec/pixel corresponds to the total field of view of about ${6.5\times6.5}$  
arcmin. The gain and readout noise were about 9.4~e$^{-}$/ADU and 13.7~e$^{-}$, 
respectively. More details about the instrumental system can be found in  
Udalski and Pych (1992). 

Observations were performed through the Cousins {\it I} filter with exposure 
times of 10~sec and sequence of ${3\times120}$~sec and through the {\it V} 
filter with exposure time of 120~sec. The map of the observed region is shown 
in Fig.~1. 

\begin{figure}[htb]
\psfig{figure=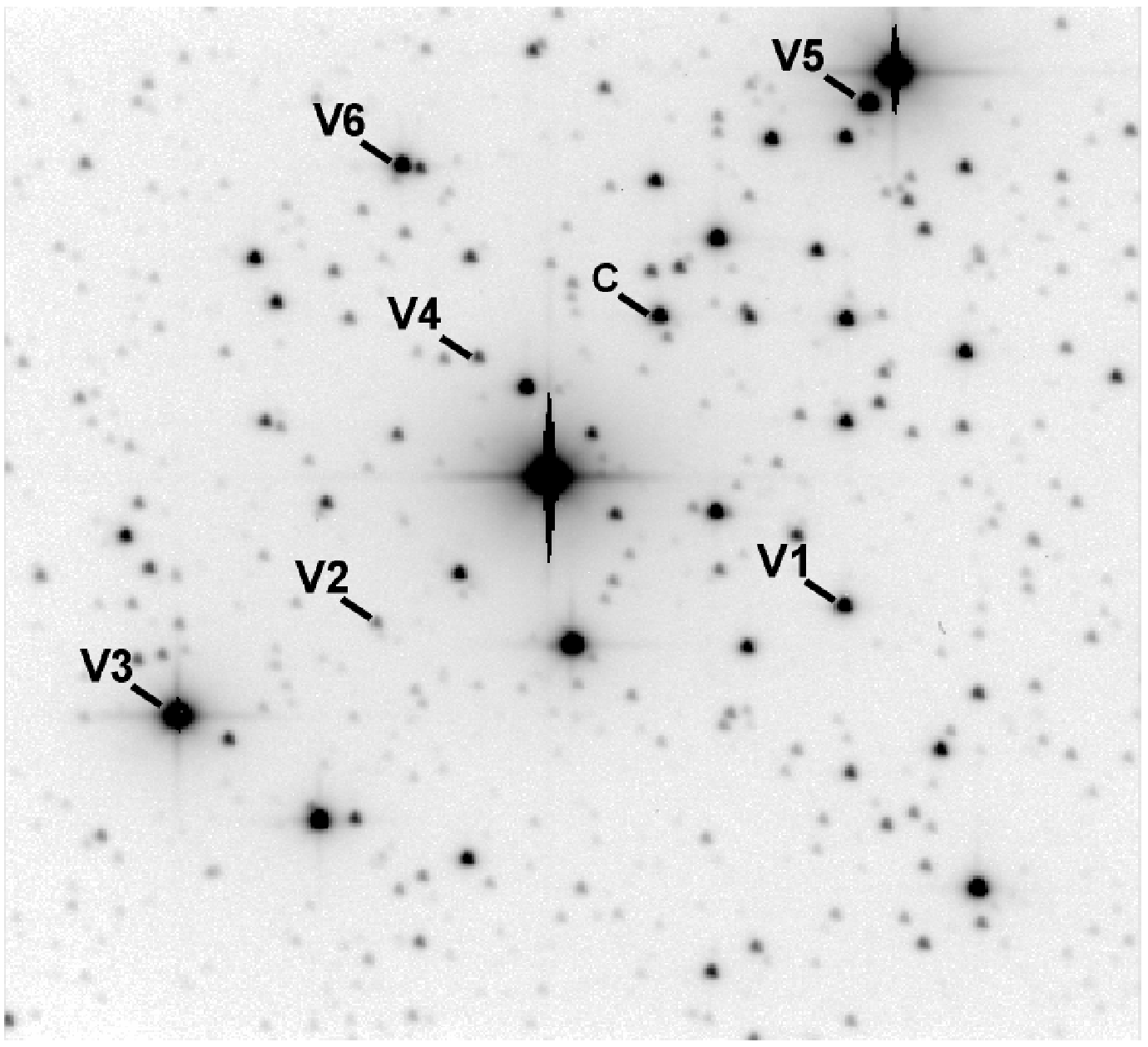,width=12.5cm}

\vskip3mm
\FigCap{Map of NGC~581 based on a frame taken with 120~sec exposure through 
the {\it I} filter. The field of view is about ${6.5\times6.5}$~arcmin. North 
is up and East to the left. V1-V6 mark detected variable stars. C points the 
comparison star.} 
\end{figure}

All images were de-biased, dark current subtracted and flat-fielded using the 
{\sc IRAF}\footnote{{\sc IRAF} is distributed by the National Optical 
Astronomy Observatories, which are operated by the Association of Universities 
for Research in Astronomy, Inc., under cooperative agreement with the NSF.} 
package. Sequences of the three 120~sec images were then co-added in the 
following way, adopting the software from the Difference Image Analysis 
(Wo{\'z}niak 2000). First, the shifts between all three frames were calculated and 
every frame was resampled into the coordinates of the first one. Then, the 
frames were summed using the IRAF package. This operation changed depth of 
our photometry from about 18~mag to about 19~mag in the {\it I} filter. 

Profile photometry was determined using {\sc Daophot} and {\sc Allstar} 
programs independently for all sets of data: 10~sec, 120~sec and ${3\times 
120}$~sec in the {\it I} filter and 120~sec in the {\it V} filter. The stellar 
point spread function (PSF) was obtained from an average of 6 visually 
selected isolated stars. Table~1 contains information on total number of 
frames, mean error returned by {\sc Daophot}, approximate brightness of 
faintest stars and number of objects in database of every set of data. 
\MakeTable{|c|c|c|c|c|}{12cm}{Information on analyzed images of NGC~581}
{
\hline
filter     & number    & {\sc Daophot}   & mag$_{\rm lim}$ & number\\
(exp.time) & of frames & error [mag]     & [mag]           & of objects\\
\hline 
$V$ (120s)          & 74  & 0.06 & 18 & 151\\
$I$ (10s)           & 231 & 0.02 & 14 & 69\\
$I$ (120s)          & 697 & 0.05 & 18 & 282\\
$I$ (3$\times$120s) & 225 & 0.04 & 19 & 325\\
\hline}

In order to create databases, images obtained at the best seeing conditions 
for every set of data were selected. Then, the master list with positions of 
all stars from the ``template'' image was prepared. Positions of stars from 
the remaining frames were compared with this list and for every star from the 
master list individual file with photometry was created. After subtraction of 
comparison star for all detected stars periodograms based on the AoV method 
(Schwarzenberg-Czerny 1989) were calculated and searched for significant 
maxima. Independently all stars were also examined visually for variability. 
This procedure was repeated for every set of data. In databases of stars 
obtained from 120~sec and combined ${3\times120}$~sec exposures made in {\it 
I} filter we found five variable stars. Another star, which was saturated on 
longer exposure frames, was found to be variable in 10~sec {\it I} filter 
database. All variables, except for the saturated one, were confirmed in the 
{\it V} filter database. 

Magnitudes of all stars were transformed from our instrumental system to 
photometry obtained by Phelps and Janes (1994), taken from the database of 
Galactic open clusters BDA (Mermilliod 1995). Common stars were identified 
visually. Then, coefficients of transformations were calculated using the 
least squares solution. Transformation equations and coefficients for summed 
{\it I} filter images and 120~sec exposure {\it V} filter are:
\begin{eqnarray*}
V&=&0.9929\cdot v-0.0632\cdot(v-i)-0.1493,\\
V-I&=&0.7874\cdot(v-i)-0.6911,\\
I&=&1.0210\cdot i+0.0939\cdot(v-i)+0.2996.
\end{eqnarray*}
Accuracy of transformations was about 0.04 mag for magnitudes in {\it V} and 
{\it I}, and about 0.03 for ${V-I}$ color. 

\Section{Variable Stars}
\begin{figure}[htb]
\centerline{\psfig{width=12.5cm,figure=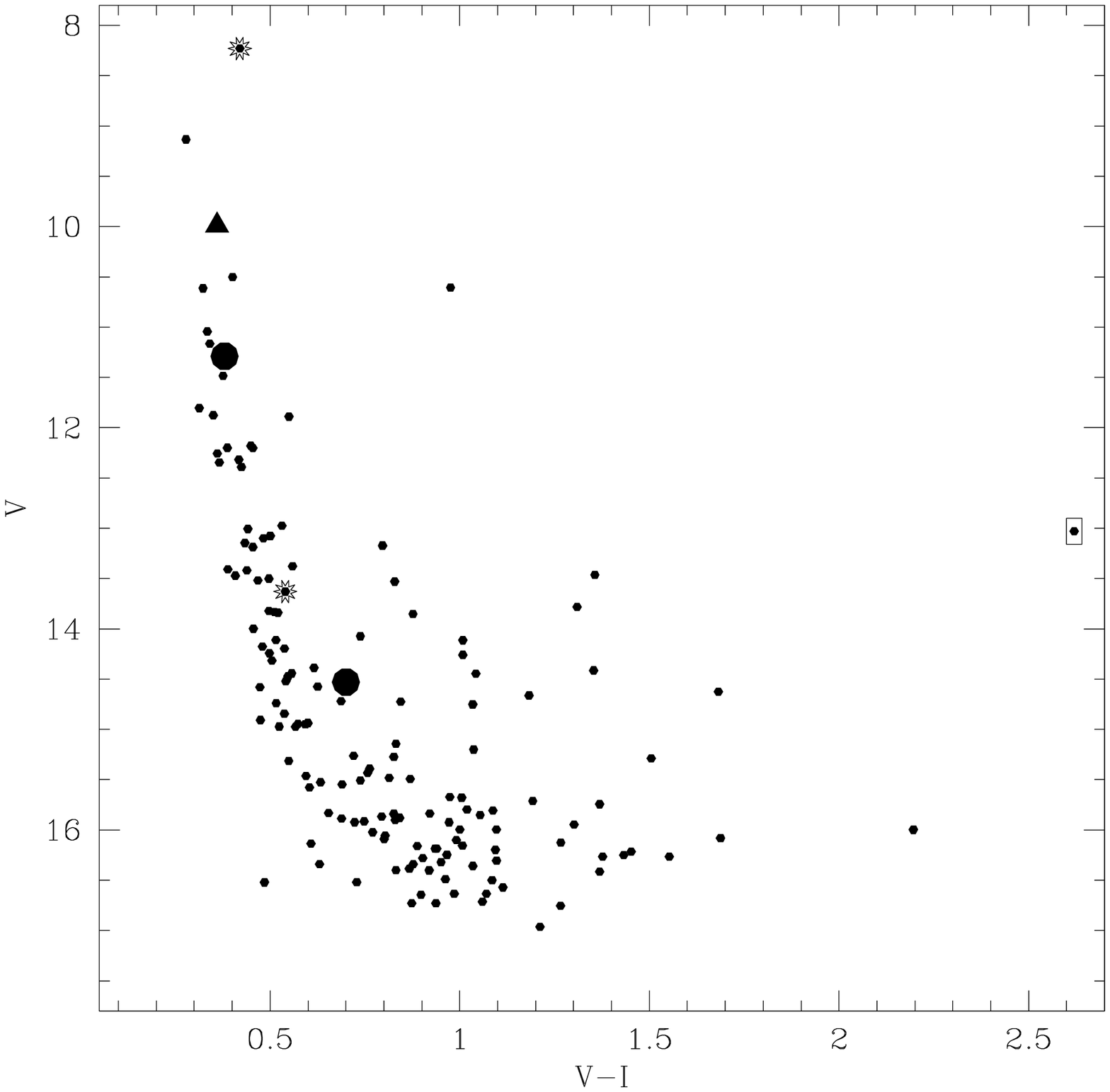}}
\FigCap{Color-magnitude diagram of the cluster NGC~581. Circles denotes 
detected eclipsing systems, triangle -- photometrically variable Be star, 
square -- pulsating red giant star. Stars mark $\gamma$~Dor candidates.} 
\end{figure}
In total six variable stars were found. Fig.~2 shows location of detected 
variable stars in {\it V} \vs {\it V--I} color-magnitude diagram (CMD). 
Table~2 contains their description. The following columns contain: 
designations of variable stars, their numbers in BDA database (Mermilliod 
1995), brightness in {\it V} and {\it I} filter (for phase 0.5 for V1 and V2, 
for HJD--2~450~000 of 1432.5 for V5 and mean brightness for V3, V4 and V6), 
times of primary minima (for V1 and V2) or maxima (for V3, V4 and V6), derived 
period (if exists) and observed amplitudes (for V1 amplitudes of primary and 
secondary minima are given). 

\MakeTable{|c|c|c|c|c|c|c|}{12cm}{Variable stars and the comparison star in 
the field of NGC~581} 
{\hline 
Star & Number & {\it V} & {\it I} & HJD          & Period & Amplitude\\
     & in BDA & [mag]   & [mag]   & -- 2450000.0 & [d]    & [mag]\\
\hline 
V1 &   42 & 11.29 & 10.91 & 1427.8223 & 4.5020? & 0.20/0.11\\
V2 &  148 & 14.53 & 13.83 & 1426.7624 & 1.6828? & 0.21\\
V3 &  144 &  8.23 &  7.81 & 1427.4674 & 0.6260  & 0.11\\    
V4 &  121 & 13.63 & 13.09 & 1427.0831 & 0.6751  & 0.11\\
V5 & 178$\dagger$& 9.99 & 9.63 & --   & --      & 0.44\\
V6 & 7089 & 13.03 & 10.41 & 1426.3096 & 1.1494  & 0.08\\
\hline 
C  & 1227 & 11.28 & 10.83 & --        & --      & --\\
\hline}

\subsection{Eclipsing Binary Stars}
Light curves of two newly discovered detached eclipsing binary systems: V1 and 
V2, are shown in Figs.~3 and~4. 
\begin{figure}[htb]  
\centerline{\psfig{width=12.5cm,figure=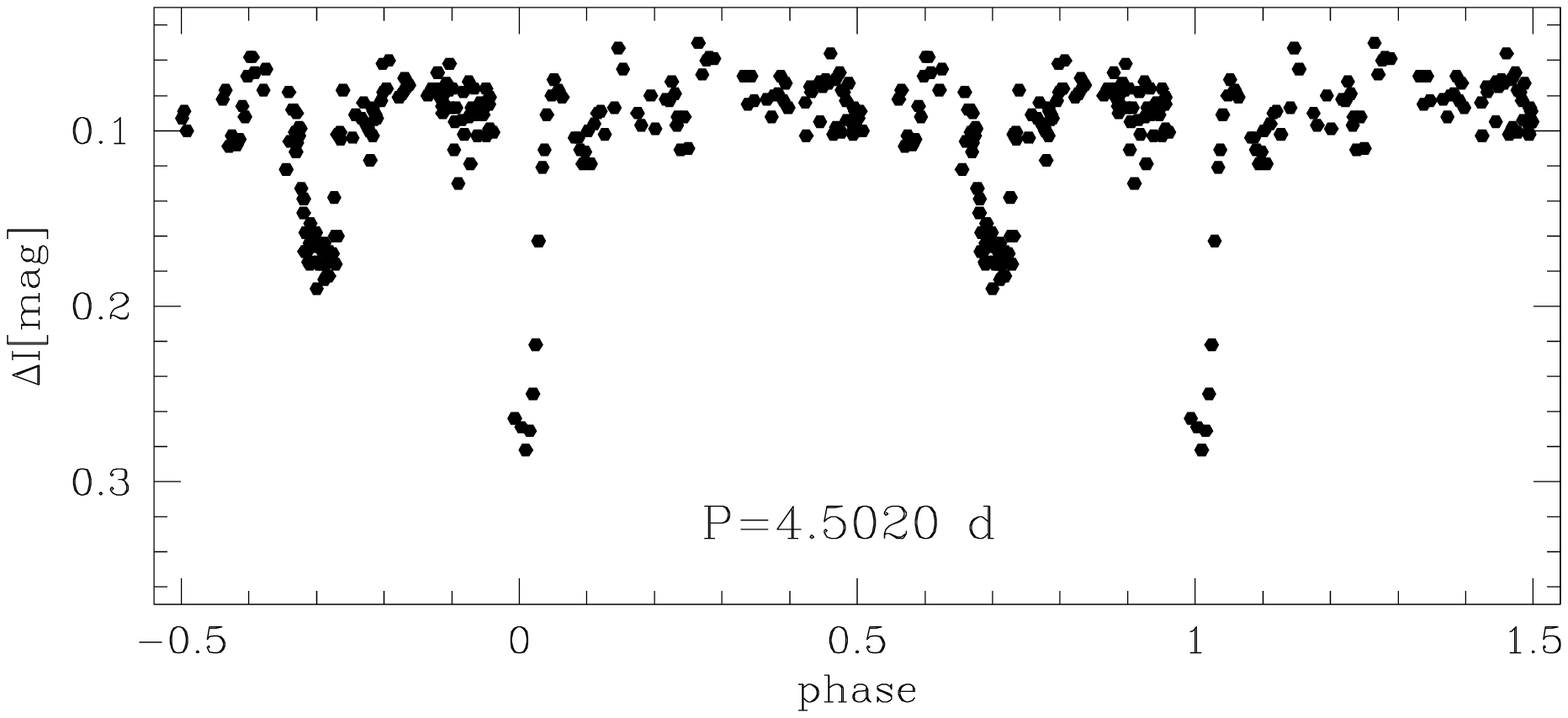}}
\FigCap{Light curve of V1. $\Delta I$ indicates the difference variable minus 
comparison.} 
\end{figure}
\begin{figure}[htb]  
\centerline{\psfig{width=12.5cm,figure=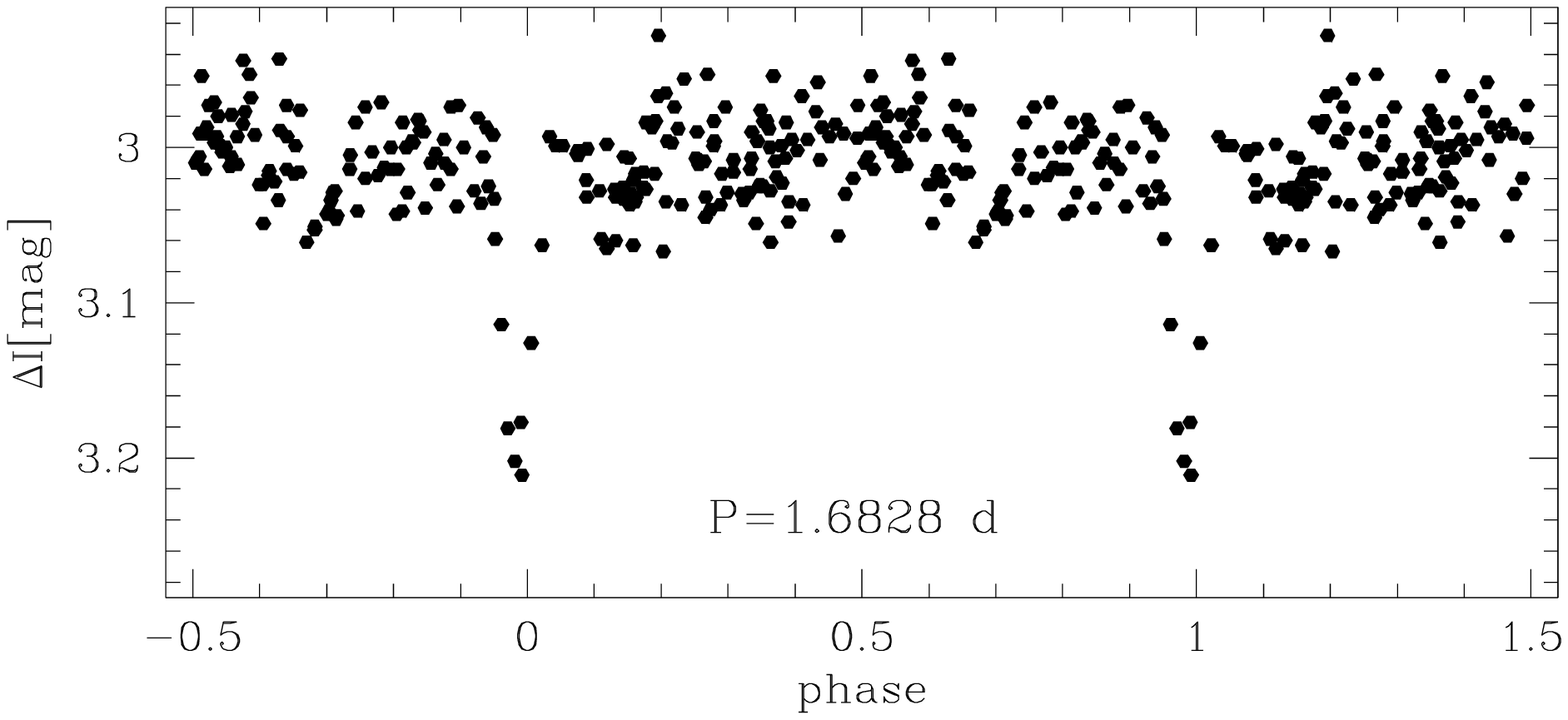}}
\FigCap{Light curve of V2. $\Delta I$ indicates the difference variable minus 
comparison.} 
\end{figure}

Light curve of variable V1, phased with the period of 4.5020~days suggests 
large eccentricity of the system orbit. Because of scarceness of our data this 
period is still preliminary. Observations can also be phased with the period 
of 6.006~days. The light curve of V1 phased with this period does not show 
eccentricity. This stars was classified by Hoag and Applequist (1965) to have 
the spectral type of B2 V, what is consistent with its position in our CMD 
(see Fig.~2). 

Period of another eclipsing binary star (V2) is also very preliminary. Some 
other periods are not excluded, but the chosen one (1.6828~d) seems to be the 
best. 

Positions of both detached eclipsing systems in the CMD (Fig.~2) support their 
cluster membership. Additionally, Sanner \etal (1999) confirms cluster 
membership of star V2 based on the study of proper motions. 

For better determination of the periods of both stars, V1 and V2, further 
observations are recommended. 

\subsection{$\gamma$~Dor Candidates}
\begin{figure}[htb]  
\centerline{\psfig{width=12.5cm,figure=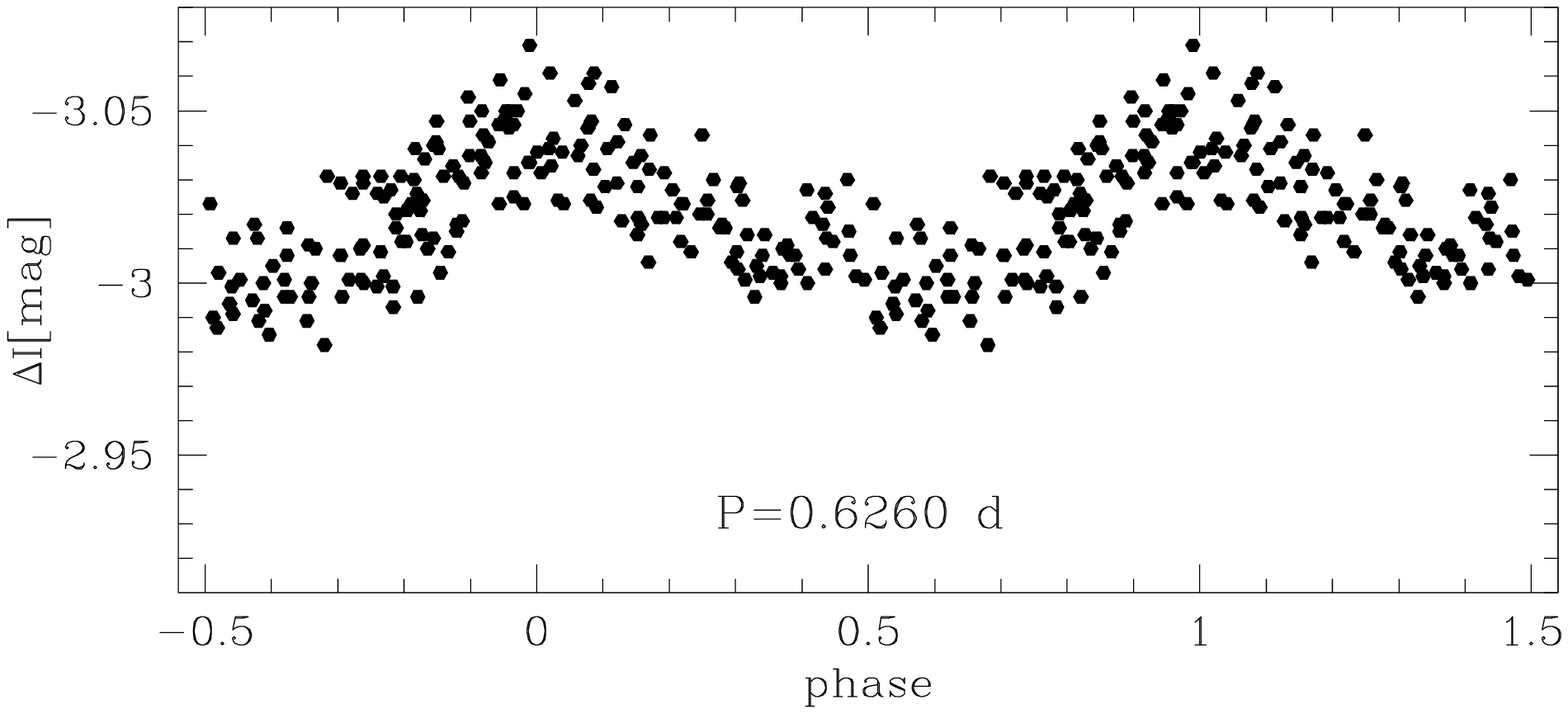}}
\FigCap{Light curve of V3. $\Delta I$ indicates the difference variable minus 
comparison.} 
\end{figure}
In Fig.~5 light curve a pulsating variable star, designated as V3 is 
presented. This star was previously investigated for its cluster membership by 
Sanner \etal (1999) (their star B). The derived proper motion turned out to be 
very different than the proper motion of the cluster. The  spectral type of V3 
is either A9~V (Sowell 1987) or F0~V (Jensen 1981). The only known pulsating  
variable stars of this type are stars of $\gamma$~Dor type (\eg Kaye \etal 
1999). The period of V3, equal to 0.6260~days, is also typical for this type 
of variable stars. Liu \etal (1989) noted variations of radial velocity of V3. 
We conclude then that star V3 is most probably a background $\gamma$~Dor star. 

\begin{figure}[htb]  
\centerline{\psfig{width=12.5cm,figure=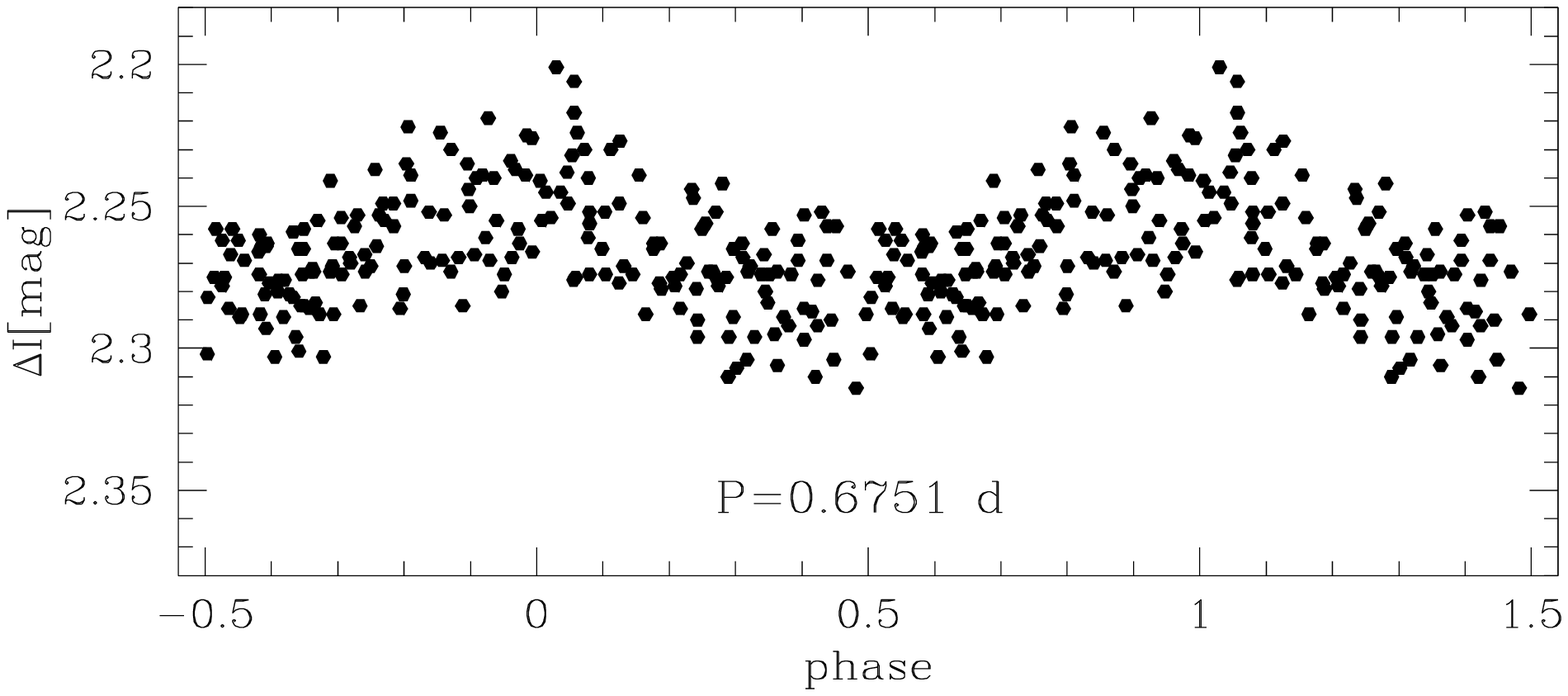}}
\FigCap{Light curve of V4. $\Delta I$ indicates the difference variable minus 
comparison.} 
\end{figure}
In Fig.~6 light curve of another $\gamma$~Dor candidate, V4 is displayed. This 
star is a member of NGC~581 (Sanner \etal 1999), what is in agreement with its 
location in the color-magnitude diagram. The most probable period of this star 
is 0.6751~days. When the same data are phased with the two times longer period 
the shape of its light curve is similar to the light curves of W~UMa stars. 
Taking into account the young age of the cluster we can exclude this 
hypothesis. Unfortunately the spectral type of this star is not available. 
Its position in the CMD suggests the spectral type of late B or early A. This 
star can be then either a Slowly Pulsating B star (SPB, Waelkens 1991) or a 
$\gamma$~Dor variable. SPB stars are known to be multi-period variables, so we 
performed check for multiperiodicity. No  other periods except for 
0.6751~days, were found. It seems therefore that this object is more likely 
another $\gamma$ Dor type star, found in the field in NGC~581. Also the 
brightness of V4 compared to other $\gamma$~Dor candidates, found in our 
previous work in NGC~659 (Pietrzy{\'n}ski \etal 2001), seems to be similar, when 
the difference of 0.5~mag in distance modulus of both clusters is taken into 
account. Spectral observations of this star would  help in distinguishing 
between these possibilities. 

\subsection{Miscellaneous Stars}
In the field of NGC~581 there are four known Be stars (Schild and Romanishin 
1976, Mermilliod 1982, Phelps and Janes 1994). Two of them were monitored 
during our observations. Star with BDA number 49, did not show  any brightness 
variability. Light curve of another Be star, designed as V5 is presented in 
Fig.~7. Our data are too scarce to allow any conclusion about possible 
periodicity of this star, but its behavior (irregular changes of the 
brightness by a couple of tenths of magnitude, over the relatively long period 
of time) is typical for  Be type star. Its position in the CMD
is consistent with the B spectral type of this star. 
Unfortunately because of the very close proximity of the very bright star (of 
${V=7.26}$~mag) and low accuracy of previously derived coordinates, it is not 
easy  to cross-identify this star with the BDA database. Most probably our 
star designated as V5 corresponds to an object with the number 178 in BDA. 
This star was discovered to be a Be star (\eg Schild and Romanishin 1976, 
Mermilliod 1982). Also the brightness and color of star 178 seems to be 
similar to V5, (\eg Burnichon 1976 gives ${V=9.90}$~mag and ${B-V= 
0.24}$~mag, Purgathofer (1964) gives ${V=10.04}$~mag and ${B-V=0.19}$~mag, 
Sagar and Joshi (1978) give ${V=10.09}$~mag, ${B-V=-0.01}$~mag, Hoag \etal 
(1961) give ${V=9.35}$~mag, ${B-V=-0.29}$~mag). One can also suspect that 
relatively large differences in brightness determinations of this star 
obtained in the past were due to intrinsic photometric variations of this 
object. 
\begin{figure}[htb]
\centerline{\psfig{width=12.5cm,figure=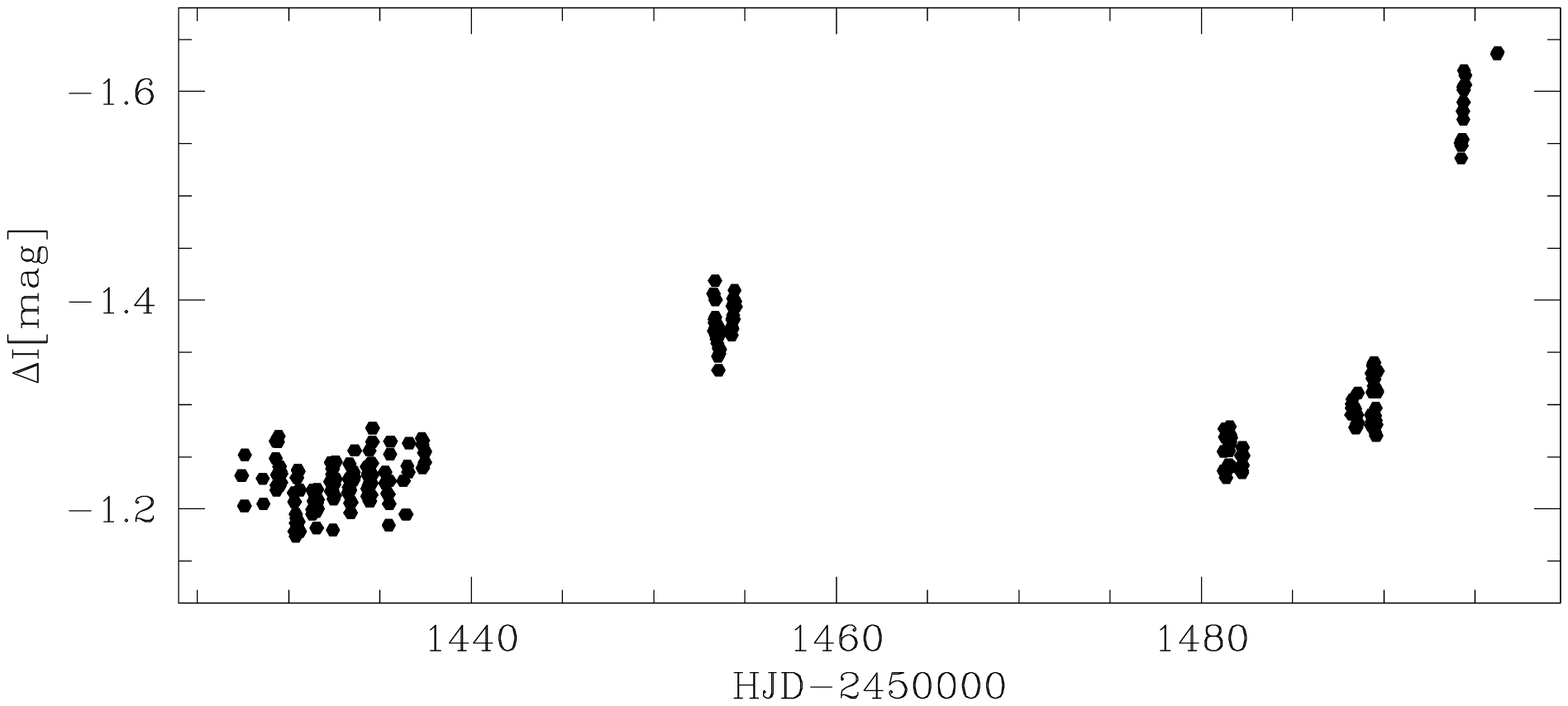}}
\FigCap{Light curve of V5. $\Delta I$ indicates the difference variable minus 
comparison.} 
\end{figure}
\begin{figure}[htb]
\centerline{\psfig{width=12.5cm,figure=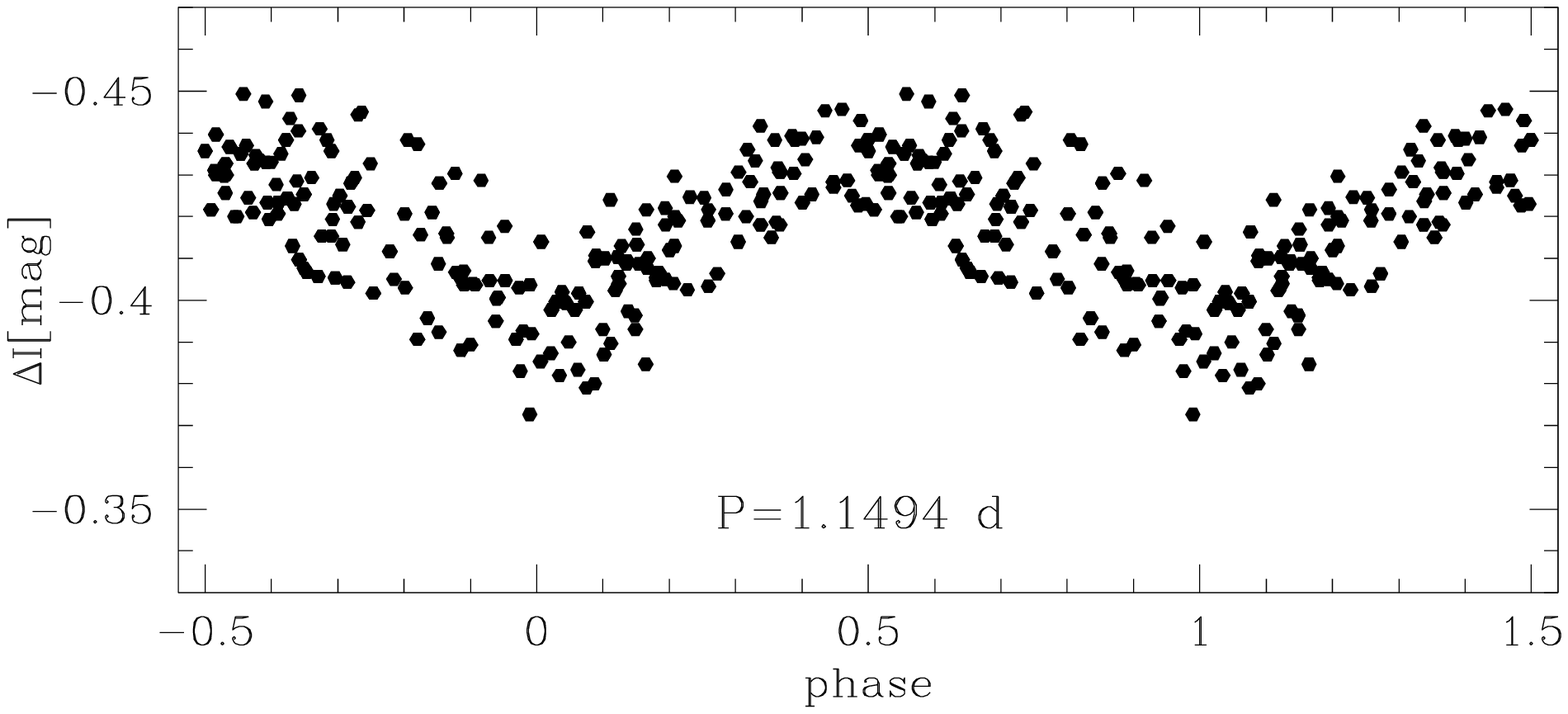}}
\FigCap{Light curve of V6. $\Delta I$ indicates the difference variable minus 
comparison.} 
\end{figure}

Light curve of another variable, V6, found in the field of NGC~581 is 
presented in Fig.~8. Brightness of this star varies  with the period of 1.1494 
days. However, the period two times longer cannot be excluded. Because V6 is very red 
with the mean ${V-I}$ color equal to 2.61~mag we suspect that, it 
is a pulsating red giant star. 

\Section{Summary}
We have presented results of a photometric search for variable stars in the 
field of young open cluster NGC~581. We have found six variable stars based on 
observations collected during 19 nights. Two detached eclipsing binaries were 
found in the field of NGC~581. Unfortunately, because of the scarceness of our 
data the periods derived for these stars are only preliminary. Stars 
designated as V3 and V4 are probably new $\gamma$~Dor candidates. Variable 
star designated as V5 seems to be the previously known Be star. Irregular 
changes of its brightness detected during our observations are typical for 
this class of stars. Another known Be star from NGC~581 was observed, but we 
did not notice any changes of brightness. Star V6 is most likely a pulsating 
red giant star. 

\Acknow{We would like to thank to Prof.\ Marcin Kubiak for his comments and 
help in preparation of this note. Thanks are also to Dr.\ Wojtek Pych and Dr.\ 
Grzegorz Pojma{\'n}ski for making their computer programs available and to 
many students for their help during observations. The work was supported by 
the KBN BW grant to the Warsaw University Observatory.}

\end{document}